\documentclass[letterpaper]{jpconf}
\usepackage{graphicx}

\begin{document}
\title{Violation of Fundamental Thermodynamic Law or Invalid Microstate Densities?}
\author{K. Michaelian$^{1}$, I. Santamar\'{i}a-Holek$^{2}$, A. P\'{e}rez-Madrid$^{3}$}

\address{ $^{1}$Inst\'{i}tuto de F\'{i}sica, Universidad Nacional Aut\'{o}noma de
M\'{e}xico,\\
A. P. 20-364, 01000 M\'{e}xico D.F., Mexico\\
$^{2}$Departamento de F\'{i}sica, Facultad de Ciencias, Circuito Exterior de
Cd. Universitario, \\
Universidad Nacional Aut\'{o}noma de M\'{e}xico, M\'{e}xico D.F.,\\
04510 Mexico\\
$^{3}$Dept. de F\'{i}sica Fonamental, Fc. de F\'{i}sica, Universitat de
Barcelona,\\
Av. Diagonal 647, 08028, Barcelona, Spain}

\ead{karo@fisica.unam.mx}
\date{\today \ PACS:\ 64.60.De, 64.60.-i,82.60.Qr}
%\maketitle

\begin{abstract}
It is often incorrectly assumed that the number of microstates $\Omega
(E,V,N,...)$ available to an isolated system can have arbitrary dependence
on the extensive variables $E,V,N,$ .... However, this is not the case for
natural systems which can reach thermodynamic equilibrium since restrictions
exist arising from the underlying equilibrium axioms of independence and 
\textit{a priori} equal probability of microstate, and the fundamental
constants of Nature. Here we derive a concise formula specifying the
condition on $\Omega $ which must be met for real systems. Models which do not respect this condition will present inconsistencies when treated under equilibrium
thermodynamic formalism. This has relevance to a number of recent models in
which negative heat capacity and violation of fundamental thermodynamic law
have been reported. Natural quantum systems obey the axioms and abide by the
fundamental constants, and thus natural systems, in the absence of infinite
range forces, can, in principle, attain thermodynamic equilibrium.
\end{abstract}

\section{Introduction}

A large number of theoretical models and experiments, purporting to
represent real systems in thermodynamic equilibrium, have been published in
which exotic thermodynamic behavior is found, such as negative
susceptibilities (e.g. negative heat capacity). These results are usually
obtained in systems with long-range forces (long with respect to the
physical size of the system). Examples are; gravitational systems \cite
{lyndenbell,lyndenbell2}, nanoclusters \cite
{bixon,schmidt,calvo,gobet,reyesnava}, heavy ion fragmentation \cite
{dagostino,chernomortez}, spin systems \cite{campa}, magnetically
self-confined plasmas \cite{kiessling}, and general models \cite{carignano}.
Other works have gone even further to claim violation of fundamental
thermodynamic law for such systems \cite{ramirez}. In light of these
persistent and disconcerting claims, it is prudent to review the statistical
mechanical foundation of equilibrium thermodynamic formalism in order to
examine the validity of applying such a formalism on these systems. It is, of course, well known that if equilibrium thermodynamic formalism
is applied to systems which are not in thermodynamic equilibrium,
inconsistencies will appear in the formalism \cite{callen}.

Equilibrium thermodynamic formalism is derivable from statistical mechanics
only under the fundamental axioms of statistical independence and \textit{a
priori }equal probability of the microstates. These conditions place
restrictions on the dependence of the number of microstates $\Omega $ on the
extensive variables $X:E,V,N,...$ for the system. In particular, we show
here that the increase in the number of microstates with any one of the
extensive variables $X$ is limited by the inequality $\Omega \Omega
^{^{\prime \prime }}/\Omega ^{^{\prime }2}<1,$where $\Omega ^{^{\prime
}}\equiv \partial \Omega /\partial X$ and $\Omega ^{^{\prime \prime }}\equiv
\partial ^{2}\Omega /\partial X^{2}$, and that model systems not respecting
this constraint do not have an extremum-maximum with respect to partition of
the extensive variable $X$ and thus can never be assumed to be in thermodynamic
equilibrium. Such models lack thermodynamic stability and are not representative of nature. Inconsistencies will then result when such models are treated
under equilibrium thermodynamic formalism. These inconsistencies should not
be considered as exotic new physics, but simply as inconsistencies arising
from unjustified application of the formalism.

There are a number of ways to construct models which do not satisfy the
extremum-maximum condition on $\Omega $. Specific examples of these will be
discussed in section 6. Nature, however, is constrained by fundamental
symmetries and constants in its system building. The existence of pure
quantum states, quantum tunneling, and interactions with the quantum vacuum,
means that natural systems respect {\it a priori} the equilibrium thermodynamic
axioms of microstate independence and equal probability. Natural systems also respect the fundamental constants of nature. One such fundamental constant, leading to the quantization of natural systems, is Planck's constant, $h$. We show here that, in the absence of infinite-range forces, quantization endows natural systems with a
number of microstates dependence on the extensive variables, $\Omega(X)$, which is a scale invariant power law. All power law relations for $\Omega(X)$ satisfy the above stability constraint on $\Omega $. Such natural systems can therefore, in the isolated time relaxed regime, always be treated with equilibrium thermodynamic formalism.

In the following section we discuss the statistical mechanical foundation of
equilibrium thermodynamic formalism and derive the above mentioned condition
of the dependence of the number of microstates $\Omega $ on the extensive
variables, required if the systems is to have an extremum-maximum with
respect to partition of an extensive variable, and thus the possibility of attaining thermodynamic equilibrium. In section 3 we describe
how this condition negates the possibility of negative heat capacity and
convex intruders in the entropy function for systems in thermodynamic
equilibrium. Section 4 shows how the condition on $\Omega $ may also be
violated, even for physically consistent models, if $\Omega $ is determined
through non-ergodic simulations. Such a violation, whatever the origin,
also leads to in-equivalence of results obtained in the microcanonical and
canonical ensembles. Section 5 demonstrates that natural quantum systems, in
the absence of infinite-range forces, obey {\it a priori} the condition on $\Omega 
$ and thus can, in principle, reach thermodynamic equilibrium. Section 6
identifies a number of common oversights in model building which lead to
violation of the condition on $\Omega ,$ and thus to inconsistencies in the
thermodynamic formalism and claims of exotic physics. Conclusions are given in section 7.

\section{Equilibrium: An Extremum-Maximum in the Number of Microstates}

Consider an isolated system with fixed extensive variables; total energy $E$%
, volume $V$, particle number $N,$ and any other extensive variable (to be
specified by ``...''). Assume that the system is composed of two subsystems,
1 and 2, and that each subsystem may have a different equation of state
(different dependencies of the number of microstates on the extensive
variables) as determined by $\Omega _{1}(E_{1},V_{1},N_{1},...)$ and $\Omega
_{2}(E_{2},V_{2},N_{2},...)$, but each individually is spatially
homogeneous. Assume that between subsystems there are fixed walls
impermeable to the exchange of all extensive variables except one, for
example the energy $E$, which is free to be exchanged between the two
subsystems. Assuming independence of the microstates, the total number of
microstates available to the combined system is, 
\begin{equation}
\Omega (E,V,N,...)=\int_{0}^{E}\Omega _{1}(E_{1}^{^{\prime
}},V_{1},N_{1},...)\Omega _{2}(E-E_{1}^{^{\prime
}},V_{2},N_{2},...)dE_{1}^{^{\prime }}.  \label{totalmicrostates}
\end{equation}

The further assumption of \textit{a priori }equal probability of the microstates
then implies that the time evolved, most probable macrostate of the isolated
system, named the \emph{equilibrium state}, is that macrostate specified by
the particular partition of the energy between subsystems which has the
largest number of microstates consistent with the all constraints. To
determine the most probable partition of the total fixed energy $E,$ we
first consider the contribution to the total number of microstates for a
particular energy partition ``p'' with definite values for $E_{1}$ and $%
E_{2}=E-E_{1}$,

\begin{equation}
\Omega _{p}(E_{1}:E,V,N,...)=\Omega _{1}(E_{1},V_{1},N_{1},...)\Omega
_{2}(E-E_{1},V_{2},N_{2},...)  \label{contribution}
\end{equation}
and look for an extremum-maximum of this function with respect to variation
of $E_{1}$ or $E_{2},$ with $E$ fixed. For example, in terms of $E_{1},$ the
extremum is determined by,

\[
\frac{\partial \Omega _{p}(E_{1}:E,V,N,...)}{\partial E_{1}}|_{E,V,N}=\frac{%
\partial \Omega _{1}(E_{1},V_{1},N_{1},...)}{\partial E_{1}}%
|_{V_{1},N_{1}}\cdot \Omega _{2}(E_{2},V_{2},N_{2},...) 
\]
\begin{equation}
+\frac{\partial \Omega _{2}(E_{2},V_{2},N_{2},...)}{\partial E_{2}}%
|_{V_{2},N_{2}}\frac{\partial E_{2}}{\partial E_{1}}\Omega
_{1}(E_{1},V_{1},N_{1},...)=0.  \label{extremcond0}
\end{equation}
It is easy to verify that this leads to the condition \cite{pathria}

\begin{equation}
\frac{\partial }{\partial E_{1}}\left[ \log \Omega _{1}(E_{1})\right] =\frac{%
\partial }{\partial E_{2}}\left[ \log \Omega _{2}(E_{2})\right] ,
\label{extremcond}
\end{equation}
where, for brevity, we have suppressed the dependence on the other, assumed
constrained, extensive variables. The connection with phenomenological
thermodynamic formalism is made by assigning the entropy of both subsystems
as \cite{callen}

\begin{equation}
S(E)\equiv k_{B}\log \Omega (E)  \label{BoltzmannEq}
\end{equation}
and thus the condition defining the extremum, Eq. (\ref{extremcond}), becomes

\begin{equation}
\frac{\partial S_{1}(E_{1})}{\partial E_{1}}=\frac{\partial S_{2}(E_{2})}{%
\partial E_{2}}\Rightarrow \frac{1}{T_{1}}=\frac{1}{T_{2}},
\label{EqTemps}
\end{equation}
implying equal temperatures. Note that equation (\ref{totalmicrostates}) and
definition (\ref{BoltzmannEq}) imply that the entropy of the combined system
(with the wall impermeable to energy removed) is only approximately
additive. For the entropy to be exactly additive, we would require that $%
\Omega (E,V,N,...)=\Omega _{1}(E_{1},V_{1},N_{1},...)\Omega
_{2}(E-E_{1},V_{2},N_{2},...)=\Omega _{p}(E_{1}:E,V,N,...)|_{T_{1}=T_{2}}$
which, however, is a good approximation for large systems in equilibrium
(thus giving extensivity in the thermodynamic limit) since in this case
there are extraordinarily more microstates consistent with the equilibrium
partition than with any other partition. The additivity or not of the
entropy, however, is not relevant to the issues to be discussed here.

The total energy $E$ will thus be partitioned among the subsystems such that
the resulting equilibrium macrostate, with definite values of $E_{1}$ and $%
E_{2},$ will be the most probable, that corresponding to the largest number
of microstates, and this will be the partition giving equal temperatures.
Likewise, the equilibrium partition of the volume, or of the number of
particles, with the other extensive variables constrained by impermeable
walls, is that which equalizes the pressure over temperature, $P/T$, or the
chemical potential over temperature, $\mu /T$, respectively \cite{pathria}.

Therefore, statistical mechanics, under the assumptions of independence and
equal probability of the microstates, explains the equilibrium state as that
macrostate, among all possible macrostates, with the maximum number of
microstates consistent with all constraints. The question then arises; 
\emph{Does an extremum-maximum in the number of microstates, with respect to
partition of an unconstrained extensive variable, always exist for whatever
dependency of }$\Omega $\emph{\ on the extensive variable? }

We will now show that the general answer to this question is \emph{no}, and
thus not all model systems can attain, even in principle, thermodynamic
equilibrium. The criterion, to be obtained below, on the functional form of
the dependence of $\Omega $ on the extensive variables, can thus be used to
discriminate between models which can attain thermodynamic equilibrium, and
models which cannot, even in principle, obtain thermodynamic equilibrium.

\subsection{The Extremum-maximum Condition}

We have tacitly assumed that the extremum as given by (\ref{extremcond0}) or
(\ref{extremcond}) corresponds to a maximum in the number of microstates.
This, however, must be verified for whatever system by determining the
curvature of $\Omega _{p}(X_{1}:\mathbf{X})$ evaluated at its extremum, i.e.
the sign of the second derivative of the contribution to the total number of
microstates of a particular partition, $\Omega _{p},$ with respect to the
unconstrained extensive variable $X_{1}$. For the extremum to correspond to
a maximum it is thus required that,

\begin{equation}
\frac{\partial ^{2}\Omega _{p}}{\partial X_{1}^{2}}=\frac{\partial
^{2}\Omega _{1}}{\partial X_{1}^{2}}\Omega _{2}+\frac{\partial ^{2}\Omega
_{2}}{\partial X_{2}^{2}}\Omega _{1}-2\frac{\partial \Omega _{1}}{\partial
X_{1}}\frac{\partial \Omega _{2}}{\partial X_{2}}<0,  \label{maxcond}
\end{equation}
where $X$ represents any extensive variable which is unconstrained, and
where $\Omega _{1}$, $\Omega _{2},$ and their derivatives, are to be
evaluated at the extremum.

Consider now the particular case that the two independently homogeneous
subsystems are identical. Since at the extremum the intensive variable,
corresponding to the unconstrained extensive variable $X_{1}$ (or $X_{2}$),
is homogeneous (see, for example, Eq. (\ref{EqTemps})) we must necessarily
have that $\Omega _{1}=\Omega _{2}$, $\partial \Omega _{1}/\partial
X_{1}=\partial \Omega _{2}/\partial X_{2}$, $\partial ^{2}\Omega
_{1}/\partial X_{1}^{2}=\partial ^{2}\Omega _{2}/\partial X_{2}^{2}$, etc.,
where each is to be evaluated at the extremum. The extremum-maximum
condition (\ref{maxcond}) then becomes

\[
\frac{\partial ^{2}\Omega _{p}}{\partial X_{1}^{2}}=2\frac{\partial
^{2}\Omega _{1}}{\partial X_{1}^{2}}\Omega _{1}-2\frac{\partial \Omega _{1}}{%
\partial X_{1}}\frac{\partial \Omega _{1}}{\partial X_{1}}<0, 
\]
or, in general (dropping the subscripts), giving that

\begin{equation}
\Omega \Omega ^{^{\prime \prime }}/\Omega ^{^{\prime }2}<1,
\label{microstatecond}
\end{equation}
where $\Omega ^{^{\prime \prime }}\equiv \partial ^{2}\Omega /\partial X^{2}$
and $\Omega ^{^{\prime }}\equiv \partial \Omega /\partial X$. Equation (\ref
{microstatecond}) is thus a very general relation that must be satisfied for
the number of microstates dependence on any one of the extensive variables
if the model is to be representative of a physically consistent system that
has an extremum-maximum and can thus reach thermodynamic equilibrium (see
Fig. 1(a)).

For example, while a power law dependence of the form $\Omega (E)\propto
\left( E/E^{*}\right) ^{D}$ , for some positive constant $E^{*}$, satisfies
condition (\ref{microstatecond}) for the exponent $D>0$, an exponential
dependence, $\Omega (E)\propto \exp (\left( E/E^{*}\right) ^{D})$, only
satisfies condition (\ref{microstatecond}) if $D<1.$ If $D=1$, $\Omega
^{^{\prime \prime }}\Omega /\Omega ^{^{\prime }2}=1$, and thus no extremum
exists. If $D>1$, then the extremum of (\ref{contribution}) is a minimum, 
\emph{not} a maximum, i.e. $\Omega ^{^{\prime \prime }}\Omega /\Omega
^{^{\prime }2}>1$ (see Fig. 1(b)). Functions of sums of two power laws (for
example $\Omega (E)\propto \left[ \left( E/E^{*}\right) ^{D}+\left(
E/E^{*}\right) ^{G}\right] $), with the powers greater than zero but
different, can also fail to satisfy (\ref{microstatecond}) near the energy $%
E^{*}$ . This latter case is directly related to the erroneous determination
of negative heat capacity for nanoclusters, which are often trapped in the
microcanonical ensemble (see below and \cite{michaelianEPL1}).

In summary, models which purport to represent real systems in thermodynamic
equilibrium, but which have a number of microstates (or phase space volume)
dependence on any one of the extensive variables which does not satisfy
condition (\ref{microstatecond}), can never reach thermodynamic equilibrium,
simply because there exists no macrostate that corresponds to an extremum
which is a maximum in the number of microstates. The extremum, if it exists,
is a minimum (see figure 1(b)), and thus homogeneity of the intensive
variables, e.g. temperature, pressure, or chemical potential, are the \emph{%
least} likely macrostates, and therefore are not synonymous with
thermodynamic equilibrium. These models may have a maximum in the number of
microstates, but it is not an extremum, it corresponds to the end points
where all of the unconstrained extensive variable will be found in one
singular part of the system, leading to inhomogeneous intensive variables.
For classical systems, this, in fact, would correspond to a singularity in
the unconstrained extensive variable at some point in the volume. Such
systems are inherently unstable and not representative of natural systems in
thermodynamic equilibrium.

\begin{figure}
\begin{center}
\includegraphics[width=350px,height=350px]{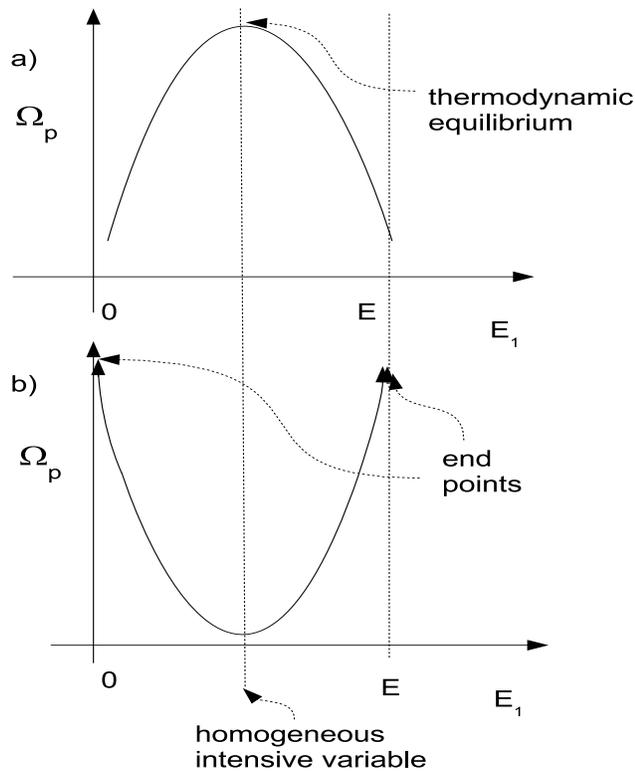}
\end{center}
\caption{\label{label}Contribution of the partition $\Omega _{p}$ to the total number of microstates as a function of the partition of the extensive variable (in
this case the energy). (a) The case of a system satisfying condition (8), a
thermodynamic equilibrium exists. (b) The case of a system not satisfying
condition (8), no thermodynamic equilibrium exists, only end points at which
all of the extensive variable is in one part of the system.}
\end{figure}

Note that for classical isolated small systems with long-range forces (with
respect to the size of the system), the factorization of the microstates,
Eq. (\ref{totalmicrostates}), will in general not apply, and the total
number of microstates of the combined system cannot be obtained by simple
decomposition of the system into subsystems. However, in this case, the very
notion of subsystems and constraints becomes unworkable. In fact, it makes
little sense to speak of the volume or the energy distribution over the
volume of such a system. The system will in general be trapped in a reduced
sub-space of the allowed phase space, the particular region being
dependent on the initial conditions. The system cannot be considered to be in
equilibrium and therefore equilibrium thermodynamic formalism simply does
not apply. If one insists on using the entropy such as the Gibbs entropy $%
S_{G}=-k_{B}\Sigma _{i=1}^{n}p_{i}\ln p_{i}$ or the Boltzmann entropy $%
S=k_{B}\ln \Omega $ (obtained from the Gibbs entropy by assigning equal
probability to each microstate $p_{i}=1/\Omega $) then one will frequently
arrive at inconsistencies such as negative heat capacity or violations of
established fundamental laws since raising the energy may un-trap the system
and allow it to reach previously unaccessible phase space (for
example, the volume, assumed constant, could suddenly increase).

Small classical systems with long range forces treated in the canonical
ensemble, however, can, in principle, be treated with equilibrium
thermodynamic formalism since in this case energy can be obtained from the
heat bath to avoid permanent trapping and the system will eventually visit
all of allowed phase space. However, the crucial point here is the word
``eventually'' since barriers in energy or angular momentum, etc., may be so
high that true thermodynamic averages become impractical for lack of time. This is the case when bimodal energy distributions are found \cite{schmidt}. For a system considered in the canonical ensemble, our equation (8) still applies but now the microstates are of the system plus heat bath.

In reality, true isolated systems do not exist in Nature. All material
exists in a world in which the system is immersed in a quantum vacuum, which
can be thought of as providing a canonical ensemble. As long as the forces
are not of infinite range, any real system therefore can in principle attain
thermodynamic equilibrium. The distrbution of the measured number of states
with any of the extensive variables for real systems (plus quantum vacuum)
must satisfy our equation (8).

\section{Negative Heat Capacity and Convex Intruders}

Using the thermodynamic definitions of the entropy, $S(E)\equiv k_{B}\ln
\Omega (E),$ the temperature $T\equiv \left( \frac{\partial E}{\partial S}%
\right) _{V,N}$ , and the heat capacity $C_{V,N}\equiv \left( \frac{\partial
E}{\partial T}\right) _{V,N}$ , it is straight forward to show that the heat
capacity in terms of the number of microstates is

\begin{equation}
C_{V,N}(E)=k_{B}\left[ 1-\Omega \Omega ^{^{\prime \prime }}/\Omega
^{^{\prime }2}\right] ^{-1}.  \label{heatcapacity}
\end{equation}
From this relation, it is obvious that if the heat capacity of a model
system is to become negative, it is required that $\Omega \Omega ^{^{\prime
\prime }}/\Omega ^{^{\prime }2}>1$, but this contradicts
the extremum-maximum condition required for a stable thermodynamic
equilibrium, Eq. (\ref{microstatecond}). Therefore, any system displaying
negative heat capacity cannot simultaneously be in thermodynamic equilibrium.

Furthermore, it is also evident from (\ref{heatcapacity}) that if the heat
capacity of a system is to change from being positive to being negative at
some energy, then it must first go through positive infinity, then flip to
negative infinity, which is obviously physically inconsistent.

Given the Boltzmann relation $S(E)\equiv k_{B}\ln \Omega (E),$ the entropy
function will show a convex intruder if 
\[
\frac{\partial ^{2}S}{\partial E^{2}}=k_{B}\left[ \frac{\Omega ^{^{\prime
\prime }}}{\Omega }-\frac{\Omega ^{^{\prime }2}}{\Omega ^{2}}\right] >0, 
\]
giving that $\Omega \Omega ^{^{\prime \prime }}/\Omega ^{^{\prime }2}>1$,
which, again, is in direct contradiction to the condition required for
thermodynamic equilibrium. Therefore, any system showing a convex intruder in the entropy function (or, equivalently, a concave intruder in the energy function)
cannot simultaneously be in thermodynamic equilibrium.

\section{Ergodicity and Inequivalence of Ensembles}

Negative heat capacity and convex intruders may result even when employing
physically consistent models but when using simulations to determine $\Omega 
$ which are not ergodic, leading to a determined $\Omega $ which does not
satisfy (\ref{microstatecond}). Long-range forces in small systems lead to
large energy barriers separating different regions of the energetically
available phase space. For classical simulations, independence and \textit{a
priori} equal probability of the microstates cannot be assumed if there is
not sufficient energy available in the system to surmount the barriers. This
occurs frequently below the solid to liquid transition energy in small
nanoclusters when treated in the classical microcanonical ensemble \cite
{michaelianEPL1}.

The \emph{determined} $\Omega $ from these simulations may not satisfy the
extremum-maximum condition (\ref{microstatecond}), even thought the
ergodically determined $\Omega $ does, since on surmounting an energy
barrier, suddenly (in energy) a large portion of previously unaccessible
phase space becomes available. The measured energy dependence of $\Omega $
may thus increase arbitrarily rapidly with energy, leading to the erroneous
determination of negative heat capacity. The problem is not easy to identify
beforehand as it is related to the particularly difficult task of proving
the ergodicity of a dynamical system.

This, however, is not an ``in principle'' problem for simulations in the
canonical ensemble since any amount of energy can be obtained from the heat
bath for surmounting whatever barrier. This assumes, however, that
sufficient time is allowed for the trajectory to visit all allowed
microstates. If not, bi-modal energy distributions at a fixed temperature
may be found \cite{michaelianEPL1}. The Laplace transform between the
ensembles is not valid for these non-ergodic simulations. This is the origin
of the inequivalence of the canonical and microcanonical ensembles for such
simulations.

In the case of convex intruders, the thermodynamically stable states
correspond to points on the tangent lines lying everywhere above the entropy
function \cite{callen}. Taking these tangent lines to define the stable
states is equivalent to ignoring the barriers and imposing equal probability
to all of the microstates, including the mixed solid-liquid states
energetically available but not accessible because of high energy barriers 
\cite{michaelianEPL1}.

Taking the determined entropy function with a convex intruder, obtained by
non-ergodic simulation as the fundamental relation, will result in
inconsistencies in the equilibrium thermodynamic formalism (see also \cite
{michaelianEPL1,michaelianEPL2,michaelianEPL3,michaelianPRL}).

\section{Natural Quantum Systems}

We now show that natural quantum systems, in the absence of infinite-range
forces, obey the equilibrium thermodynamic axioms of independence and equal
probability of the microstates and, as a result of quantization, have a
number of microstates dependence that satisfies condition (\ref
{microstatecond}) at all energies.

Non-ergodicity, or trapping by barriers, is \emph{not} an ``in principle''
problem in quantum systems because of quantum tunneling and interaction of
the system with the quantum vacuum. The principle of detailed balance,
resulting from the time reversibility of the Schr\"{o}dinger equation, then
ensures that all microstates have \textit{a priori} equal probability \cite
{callen}.

The existence of a finite Planck's constant implies that the energy, volume,
particle, and any other extensive variable, eigenstates are quantized.
Consider, for example, a state space of a simple system of fixed energy $E,$
volume $V$, and particle number $N,$ containing $D$ degrees of freedom for
storing energy. Assume that the energy is quantized in units of $
%TCIMACRO{\UNICODE[m]{0x127}}
%BeginExpansion
\rlap{\protect\rule[1.1ex]{.325em}{.1ex}}h%
%EndExpansion
\omega _{0}$ (for example, the Einstein model of a solid \cite{callen},
where $D=3N$, with $N$ equal to the number of particles). The number of
microstates available to this system corresponds to the number of ways to
distribute the $E/%TCIMACRO{\UNICODE[m]{0x127}}
%BeginExpansion
\rlap{\protect\rule[1.1ex]{.325em}{.1ex}}h%
%EndExpansion
\omega _{0}$ energy packets over $D$ degrees of freedom, which is \cite
{callen}

\begin{equation}
\Omega (E)=\frac{(D-1+E/%TCIMACRO{\UNICODE[m]{0x127}}
%BeginExpansion
\rlap{\protect\rule[1.1ex]{.325em}{.1ex}}h%
%EndExpansion
\omega _{0})!}{(D-1)!(E/%TCIMACRO{\UNICODE[m]{0x127}}
%BeginExpansion
\rlap{\protect\rule[1.1ex]{.325em}{.1ex}}h%
%EndExpansion
\omega _{0})!}\approx \frac{1}{D!}\left( \frac{E}{%
%TCIMACRO{\UNICODE[m]{0x127}}
%BeginExpansion
\rlap{\protect\rule[1.1ex]{.325em}{.1ex}}h%
%EndExpansion
\omega _{0}}\right) ^{D}=\frac{1}{3N!}\left( \frac{E}{%
%TCIMACRO{\UNICODE[m]{0x127}}
%BeginExpansion
\rlap{\protect\rule[1.1ex]{.325em}{.1ex}}h%
%EndExpansion
\omega _{0}}\right) ^{3N},  \label{einsteinsolid}
\end{equation}
where the result to the right of the ``$\approx $'' sign is obtained by
calculating the number of quantum states with energy less than, or equal to, 
$E,$ which is a very good approximation for the number of microstates $%
\Omega (E)$ with energy between $E$ and $E+\delta E,$ as long as $D\gg 1$ 
\cite{callen}. The number of microstates of a quantum system thus increases
as a scale invariant power law in energy $E,$ with the exponent of the power
law being equal to the number of independent degrees of freedom $D$. This
result does not depend on whether the frequencies of all modes are the same, 
$\omega _{0},$ or whether there exists a spectrum of values $\omega _{i}$.

A similar power law can be obtained for the number of microstates dependence
on the other extensive variables, such as the volume $V,$ or the particle
number $N$. For example, the dependence of $\Omega $ on the number of
particles $N$ can be found by holding constant the total energy and allowing 
$N$ to vary. By the inherent symmetry apparent in the term to the left of
the ``$\approx $'' sign in equation (\ref{einsteinsolid}), and for $E/\hbar
\omega _{0}\gg 1,$ it is easy to see that the result is

\[
\Omega (N)\approx \frac{1}{\left( \frac{E}{\hbar \omega _{0}}\right) !}%
\left( 3N\right) ^{E/\hbar \omega _{0}}. 
\]
It is trivial to show that all power law relations satisfy condition (\ref{microstatecond}), thus (\ref{microstatecond}) is satisfied by natural quantum systems at all energies, particle numbers, and volumes. Therefore, all isolated natural systems in the absence of infinite-range forces can, in principle, attain stable
thermodynamic equilibrium.

In the limit $\hbar \rightarrow 0,$ the exponent of the power law $%
\rightarrow \infty $, and it is easy to verify that $\Omega \Omega
^{^{\prime \prime }}/\Omega ^{^{\prime }2}\rightarrow 1,$ and thus, in this
limit, there exists no extremum-maximum, no thermodynamic equilibrium.
Quantization of the energy, or the existence of a finite Planck's constant,
is therefore a necessary condition to ensure thermodynamic stability of an
isolated macroscopic system. The same holds true for quantization of
particle number, volume, and spin, etc. This is the macroscopic analogue of
the fact that quantization is a necessary condition to ensure the stability
of the atom.

\section{Non-physical Model Design}

Models meant to provide insight on the thermodynamic behavior of natural
systems must have incorporated in their construction all the degrees of
freedom and their correspondent constraints that the system is subject to in
Nature. For example, models representing isolated systems must respect
conservation of energy, momentum, angular momentum, volume, charge, particle
number, etc. However, there are other constraints that arise because of the
fundamental constants of Nature. For example, maximum signal rates (velocity
of interaction) cannot exceed the velocity of light, $c$. Also, the number
of microstates within a given volume $\Gamma $ of state space cannot exceed
the value $\Gamma /\hbar ^{D/2}$,\cite{callen} where $D$ is the
dimensionality of state space, which is limited by the finite value of
Planck's constant $h.$ This requirement is equivalent to the quantization of
thermodynamic variables, including those arising from internal degrees of
freedom, such as spin. Furthermore, quantum tunneling and interaction with
the vacuum, plus the reversibility of the Schr\"{o}dinger equation, then
ensures equal probability of the microstates \cite{callen}. In counting
microstates, it is also important to take into account the
indistinguishability of elementary particles and their particular statistics
(Fermi-Dirac, or Bose-Einstein). Although at high temperatures or low
pressures, the quantum description often coincides with the classical
description for some thermodynamic properties, this is not the case for
counting microstates, as the Gibb's paradox makes explicitly clear \cite
{callen}.

Relevant models must also be physically consistent. For example, internal
interaction fields providing an exchange of energy cannot be presumed to
exist in isolation, but must necessarily arise from internal material
sources. External fields affecting the internal energy or entropy, must be
taken into account by, for example, modifying the chemical potential \cite
{landau}.

Still another important condition for the thermodynamic analysis to be free
of inconsistencies is that the number of extensive variables used in the
analysis must be commensurate with the number of degrees of freedom in the
models state space, which depends on the particular Hamiltonian of the
system \cite{callen,jaynes}. For example, inconsistencies in the
thermodynamic analysis of molecular hydrogen gas led to the discovery of
ortho- and para-hydrogen \cite{callen}.

Models not respecting these restrictions will present inconsistencies when
treated under equilibrium thermodynamic formalism constructed on the basis
of the existence of these \cite{callen}. In particular, such models may fail
the condition for the existence of a extremum-maximum, $\Omega \Omega
^{^{\prime \prime }}/\Omega ^{^{\prime }2}<1.$ Neglect of these
considerations when constructing models has often lead to the reporting of
the inconsistencies as if they were indicative of new physical phenomena, or
of violations of fundamental thermodynamic law. Below we identify particular
cases exemplifying some of the typical neglect of consideration mentioned above.

In \cite{bixon}, a model is presented purporting to demonstrate the
possibility of negative heat capacity in nanoclusters. The energies of the
microstates in this model are arbitrarily assigned and not derived from a
physically consistent Hamiltonian using the Schr\"{o}dinger equation, or by
quantization of a classical phase space volume. The model, in fact, violates the
condition of energy quantization and has an arbitrarily large degeneracy in
energy of the higher energy states, leading to a violation of the
extremum-maximum condition on $\Omega $. Such models do not respect a finite
Planck's constant which implies a power law distribution of the energy states, and furthermore do not respect correct particle
statistics, and, therefore, cannot be used for elucidating equilibrium
thermodynamic properties of real nanoclusters.

Another model, also purporting to lend support for negative heat capacity in
real systems, is given in \cite{lyndenbell2}. Here, internal fields not
arising from material sources are included, and the volume is arbitrarily
defined by a non-interacting sphere. Alternatively, if the authors had
intended the field to be external, then it should have been taken into
account consistently by defining new intensive variables (e.g. new chemical
potentials) that includes the energy exchanged with the system \cite
{michaelianEPL2}.

Non-ergodic experiments or simulations, although employing physically
consistent models, can lead to a determined energy dependence of $\Omega $
that does not respect the extremum-maximum condition, thereby giving
``apparent'' negative heat capacity. Such results on nanoclusters have been
frequently presented in the literature \cite
{bixon,calvo,gobet,reyesnava,schmidt,briant,wales}. The prevailing consensus
regarding these results is that negative heat capacity arises from a
particular, but inherent, distribution of the energy states of the
nanocluster \cite{bixon,calvo,schmidt,wales}. However, these explanations
consistently ignore the additional degrees of freedom for carrying
energy that arise as the system changes phase. For example, proponents of
this explanation derive an identical heat capacity for the liquid as
for the solid, which is clearly in contradiction with both experiment and results
from simulations \cite{reyesnava}. It is, in fact, the
occulting of these additional degrees of freedom in the microcanonical
ensemble due to large energy barriers that leads to the erroneous
determination of negative heat capacity \cite{michaelianEPL1}.

The correct $\Omega $ for nanoclusters contains sums of
contributions of the different mixed solid-liquid states, and these
contributions increase with energy as different power laws \cite
{michaelianEPL1}. The powers are different because of the different number
of degrees of freedom for storing energy for those atoms in the liquid
portion compared to those in the solid portion. In the liquid phase, in
addition to the vibrational modes, the atoms can carry kinetic energy
related to their diffusion about the cluster. This gives rise to the
observed different heat capacities of the two phases. Including the mixed
solid-liquid states at the relevant energy results in no negative heat
capacity being found \cite{michaelianEPL1}. Including only the states in which all atoms are in the solid, and then, on passing a high barrier at energy $E^{*}$, including only the states in which all atoms are in the liquid, gives a determined number of microstates with an energy
dependence of form $\Omega (E)\propto \left[ \left( E/E^{*}\right)
^{D}+\left( E/E^{*}\right) ^{G}\right] $) where $D$ is proportional to the
number of degrees of freedom for storing energy in the solid and $G$ is
proportional to the number of similar degrees of freedom in the liquid. As mentioned
in section 2.1 this functional form for $\Omega $ can, depending on the
difference of $G$ from $D$ (which also depends on the size of the system),
violate the extremum-maximum condition (\ref{microstatecond}) and thus lead
to apparent negative heat capacity.

Still another example of violating the condition on $\Omega ,$ concluding
negative heat capacity, and further suggesting the violation of a
fundamental thermodynamic law, is given in \cite{ramirez}. The model
consists of $N$ rotors (spin). A kinetic energy term and an infinite-range
spin-spin interaction are modeled by a Hamiltonian of form $%
\sum_{i=1}^{N}p_{i}^{2}/2+J/N\sum_{i,j=1}^{N}(1-\cos (\theta _{i}-\theta
_{j})),$ where $J$ is the spin-spin coupling constant, and $\theta $ the
angle of the rotor with respect to a given axis$.$ This system, in fact, has
no spatial dimension so the Hamiltonian does not depend on the volume of the
system. When the thermodynamic analysis of this system is carried out using
an entropy function dependent on only two extensive variables, the total
energy per particle and a total magnetization order parameter $m$, the
thermodynamic results are consistent, and the microcanonical and canonical
ensemble results are equivalent \cite{ramirez}.

However, the authors then add a ``nearest neighbor'' interaction to the
Hamiltonian of form $-K\sum_{i=1}^{N}\cos (\theta _{i+1}-\theta _{i})$, with
periodic boundary conditions$.$ The introduction of a nearest neighbor
interaction implies the introduction of a new coordinate. The most
reasonable interpretation of their model leads us to suppose that this
coordinate is one-dimensional \emph{length}. Spin density, and also energy
density, then become relevant intensive variables to be distributed over
this new coordinate. Because of the symmetry inherent in such a one
dimensional chain of rotors, one would normally expect a macroscopically
homogeneous distribution of the magnetization and energy for such a system
in thermodynamic equilibrium. However, the authors find that the
distribution of these variables is non-homogeneous when the entropy is
maximum, implying a situation similar to that of figure 1(b). Their system
does not have an extremum-maximum, but instead endpoint maxima at which all
the magnetization or energy is in one part of the system.

According to our analysis given in the previous sections, this results from
a number of microstates, as a function of the partition of the energy or the
magnetization, which fails to satisfy the condition for the existence of an
extremum-maximum, Eq. (\ref{microstatecond}). This failure, in turn, is
related to the non-physical nature of the model (infinite-range interaction
and non-quantization of spin nor of the energy, see section 5). Negative
heat capacity and ``evidence'' for the violation of the zeroth law of
thermodynamics are simply inconsistencies obtained on a non-physical model
which is not in thermodynamic equilibrium and which, in fact, can never
reach thermodynamic equilibrium.

\section{Conclusions}

Using a model to obtain the thermodynamic properties of a real system requires recognition of all the macroscopic and internal degrees of freedom as well as their
constraints. We have identified a general condition (\ref{microstatecond})
on the dependence of the number of microstates on any of the extensive
variables which must be satisfied if the model system is to have the possibility
of attaining thermodynamic equilibrium. Model systems not satisfying this
condition are systems which can never reach thermodynamic equilibrium.
Negative heat capacity, convex intruders in the entropy function,
non-physical singularities, differences obtained in different ensembles, and
violations of fundamental thermodynamic laws are all inconsistencies
resulting from the application of equilibrium thermodynamic formalism to
these deficient model systems. The quantization inherent in natural systems ensures the compliance with condition (\ref{microstatecond}) giving an extremum-maximum
in the number of microstates with respect to partition of an unconstrained
extensive variable, and thus the possibility of natural systems to arrive at thermodynamic equilibrium.

%\section{Acknowledgments}
\ack
The authors are grateful for comments on the manuscript by J.
Gonz\'{a}lez-Gonz\'{a}lez, and for the financial support of DGAPA-UNAM,
projects IN-102609 and IN103113.
\\
\\
\noindent
%\section*{References}
%\begin{thereferences}

\end{document}